\documentclass[%
 reprint,
 amsmath,amssymb,
 aps,
floatfix,
]{revtex4-2}

\usepackage{graphicx}
\usepackage{dcolumn}
\usepackage{bm}
\usepackage{scalerel}
\usepackage{textcomp}
\usepackage{amsmath}
\usepackage{physics}
\usepackage[caption=false]{subfig}
\usepackage{booktabs}

\usepackage{xcolor}
\definecolor{darkblue}{RGB}{0,0,150}
\definecolor{nightblue}{RGB}{0,0,100}
\usepackage[
bookmarksnumbered,
pdfpagelabels=false,
colorlinks,
linkcolor=nightblue,
citecolor=nightblue,
filecolor=black,
urlcolor=darkblue,
breaklinks=true
]{hyperref}
\begin{document}

\preprint{APS/123-QED}

\title{Theory of Multi-Orbital Topological Superconductivity\\in Transition Metal Dichalcogenides}

\author{Gilad Margalit}
\author{Erez Berg}
\author{Yuval Oreg}
 \affiliation{Department of Condensed Matter, Weizmann Institute of Science.}

\date{\today}

\begin{abstract}
We study possible superconducting states in transition metal dichalcogenide (TMD) monolayers, assuming an on-site pairing potential that includes both intra- and inter-orbital terms. We find that if the mirror symmetry with respect to the system's plane is broken (e.g., by a substrate), this type of pairing can give rise to unconventional superconductivity, including time-reversal-invariant nodal and fully gapped topological phases. Using a multi-orbital renormalization group procedure, we show how these phases may result from the interplay between the local Coulomb repulsion, Hund's rule coupling, and phonon-mediated attraction. In particular, for a range of interaction parameters, the system transitions from a trivial phase to a nodal phase and finally to a gapped topological phase upon increasing the strength of the mirror symmetry breaking term.



\end{abstract}

\maketitle

\section{Introduction}
\label{intro}

Time-reversal-invariant phases of topological superconductivity have long been theorized \cite{Schnyder2008, Qi2009}, yet observation of topologically-protected edge states in superconductors has remained scarce. An analogous phase arises in superfluid $^3$He-B~\cite{Volovik2003,Vollhardt2003}, which realizes a Hamiltonian in the same class as a time-reversal-invariant topological superconductor (TRITOPS)~\cite{Chung2009, Qi2011, Haim2019} in 3D. 
Several realizations of a TRITOPS phase have been proposed in solid state systems, either intrinsically~\cite{Fu2010,Scheurer2015} or in engineered systems of semiconductors proximity-coupled to superconductors~\cite{Zhang2013b,Berg2013,Haim2014}.

Related to this phase are time-reversal-invariant nodal superconducting phases, which in 2D involve the bulk superconducting gap closing at specific points. Whereas gapped TRITOPS phases support topologically protected modes on all edges, nodal superconducting phases typically exhibit zero-energy states on edges in certain directions~\cite{Hu1994,Tanaka1995,Kashiwaya1995,Deutscher2005,Lee2012}. Gap nodes have been observed in unconventional superconductors, such as cuprates and certain iron-based superconductors. They have also been theorized in transition metal dichalcogenide (TMD) systems~\cite{Hsu2017}.

In this work, we propose a mechanism to realize both nodal and gapped time-reversal-invariant topological superconductivity from the interplay between Coulomb repulsion and phonon-mediated attraction in multi-orbital systems. 
Though models of superconductivity often neglect the effect of multi-orbital physics, implicitly assuming that pairing only takes place between states of the same orbital, this need not be the case. Phonons can couple any two electron states with opposite momenta which preserve the discrete symmetries of the Hamiltonian. Thus, one typically finds several competing pairing states in a multi-orbital system. The relative strength of the pairing channels determines which of the distinct superconducting phases is realized.

As a concrete example of this, we examine the inter-orbital pairing present in a doped 2D 1H-TMD monolayer of tantalum disulfide (TaS$_2$) where a substrate breaks mirror symmetry with respect to the plane. 
Restricting our scope to only on-site pairing interactions, we find that multiple intra- and inter-orbital terms are allowed by symmetry. 
We show that increasing the magnitude of a particular inter-orbital pairing term drives the system from a conventional superconductor to a nodal SC, and then to a fully gapped TRITOPS.


Recently, edge modes were observed on the surface of superconducting 4Hb-TaS$_2$~\cite{Beidenkopf2021}. In that work, we theorize that these edge modes are due to a multi-orbital nodal phase. 
We expand this analysis here, deriving the phase diagram starting from a microscopic model with an on-site Coulomb repulsion, Hund's rule coupling, and phonon-mediated attraction between electrons. 
Using a multidimensional renormalization group (RG) method, we find conditions under which multi-orbital topological superconductivity occurs. 
In particular, in the TaS$_2$ case, we find that this requires the inter-orbital electron-phonon coupling strength to be larger than the intra-orbital. This type of RG analysis will be helpful for finding future candidate materials exhibiting multi-orbital pairing.

This paper is organized as follows. In Sec. \ref{model}, we outline a tight-binding model of a 1H-TaS$_2$ monolayer. In Sec. \ref{sc}, we add superconductivity to the model, allowing all on-site pairing terms that do not violate the symmetries of the lattice. We show that certain ranges of pairing values can give rise to a nodal phase. We then explicitly add Rashba-type spin-orbit coupling (SOC) in Sec. \ref{tritops}, which enhances the separation between the nodes and allows for a transition into a gapped TRITOPS phase. A topological invariant is computed and verified to be nontrivial in Sec. \ref{topinv}. Finally, after describing the experiment in which this type of multi-orbital nodal SC may have been observed (Sec. \ref{exp}), we implement our RG method in Sec. \ref{rg}.

\section{Model and Symmetries}
\label{model}

\begin{figure}

     \subfloat[]{\includegraphics[width=0.23\textwidth]{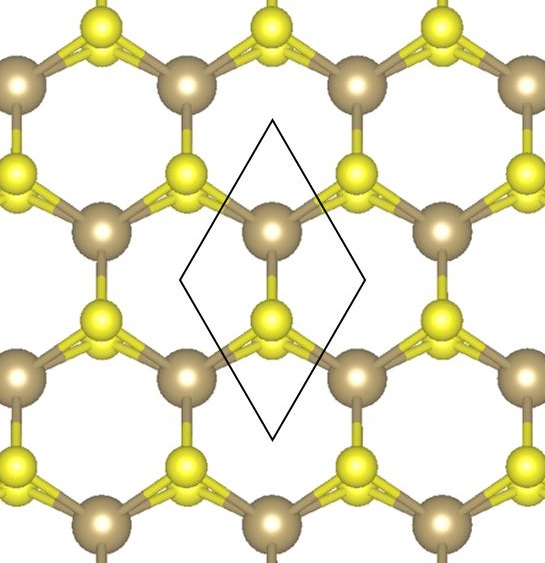}
        \label{vesta}}
     \subfloat[]{\includegraphics[width=0.23\textwidth]{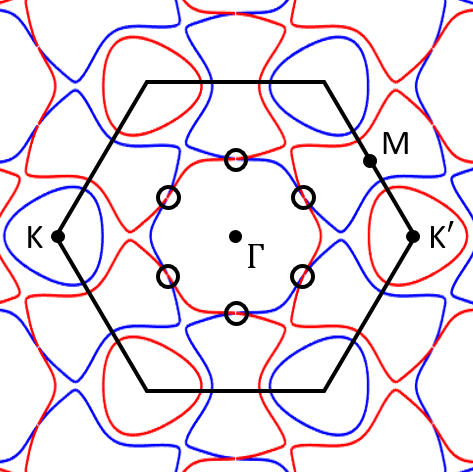}
        \label{fermi}}
        
    \subfloat[]{\includegraphics[width=0.45\textwidth]{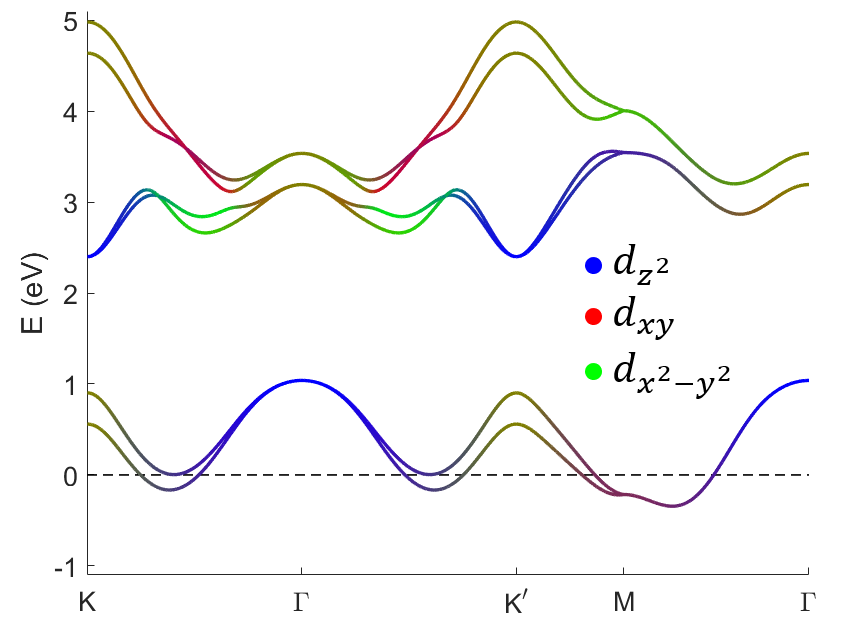}
        \label{bands}}
        
        \caption{(a) Crystal structure of the 1H-TaS$_2$ monolayer. Tantalum atoms (green) lie in the monolayer plane, while Sulfur atoms (yellow) extend symmetrically above and below the plane. The unit cell is marked in black. (b) Fermi surface of the TaS$_2$ model in the absence of Rashba SOC. The hexagon, points, and circles indicate the Brillouin zone, high-symmetry points, and band-crossings, respectively. Color shows the spin degree of freedom; due to Ising SOC, states at opposite momentum are required to have opposite spin. (c) Band structure of fitted TaS$_2$ tight-binding model. Two bands cross the Fermi energy (dotted line). The orbital composition of each state is indicated by its color.}
\end{figure}

We model a 2D monolayer of 1H-TaS$_2$ on a substrate. The monolayer crystal structure is shown in Fig.~\ref{vesta}. In the recent experiment that inspired this study~\cite{Beidenkopf2021}, the 1H-TaS$_2$ is the top layer of a slab of 4Hb-TaS$_2$, although the model applies more generally. The presence of the substrate is crucial because it breaks $M_z$, the mirror symmetry along the $z$-direction perpendicular to the monolayer plane. The van der Waals attraction between the monolayer and the substrate is weak enough that we can treat the system as 2D.

The single-particle part of the Hamiltonian can be well-described by a tight-binding model with up to third-nearest-neighbor hopping \cite{Law2018, Khodas2018}, defined in detail in Appendix \ref{tb}. The hopping matrices are constrained by the symmetries of the TMD monolayer: rotation by 120 degrees about the z-axis through a Ta site ($C_3$), reflection along the $x$-axis about a Ta site ($M_x$), time reversal ($T$), and reflection along the $z$-axis ($M_z$) (the latter symmetry is broken by the substrate, whose effects we will discuss in Sec. \ref{tritops}). The model parameters are obtained by fitting the band structure to a Density Functional Theory (DFT) calculation of an isolated monolayer \footnote{DFT results are obtained using the VASP package with Generalized Gradient Approximation (GGA) functionals~\cite{GGA}.}. The chemical potential is then shifted by -50 meV with respect to the monolayer DFT fit in order to align the modeled Fermi surface (Fig. \ref{fermi}) with ARPES measurements taken of the surface of 4Hb-TaS$_2$ \cite{Kanigel2020}.

Our model consists of a triangular lattice of positive Ta ions, with three relevant orbitals per ion: $d_{z^2}$, $d_{xy}$, and $d_{x^2-y^2}$. Including the spin degree of freedom, we have 6 states per site, yielding a 6-band model. This accurately reflects the \textit{ab initio} results, since the calculated states have a negligible contribution from the Ta $p$ orbitals, while the other two $d$ orbitals are far from the Fermi level due to crystal field splitting.

Spin remains a good quantum number for the isolated layer despite significant intrinsic SOC. Since this SOC originates within the plane, it restricts the eigenstates to have spins oriented out-of-plane, with states at opposite momentum having opposite spin. We call this type of SOC ``Ising SOC" to distinguish it from Rasha SOC, which is imposed externally on the monolayer when $M_z$ is broken. Rashba SOC will be added to the model in Sec. \ref{tritops}.

After fitting the band structure, we obtain a single-particle Hamiltonian $H_0(\bm{k})$ as a function of momentum $\bm{k}$. Its band structure is plotted in Fig. \ref{bands}.

\section{Superconductivity}
\label{sc}

\subsection{Pairing Matrix}
\label{pair}

We extend the bare Hamiltonian $H_0(\bm{k})$ from the previous section into an effective superconducting Hamiltonian $H_{SC}(\bm{k})$ by means of the Boguliubov-de-Gennes (BdG) formalism. The resulting Hamiltonian has the form
\begin{equation} \label{Hsc}
H_{\rm SC}(\bm{k}) =\left[\begin{array}{cc}
H_0(\bm{k}) & \Delta\\
\Delta^{\dagger} & -H_0^{*}(-\bm{k})
\end{array}\right].
\end{equation}
The Hamiltonian acts on the Nambu vector $\Psi^\dagger_{\bm{k}} = (\psi^\dagger_{\bm{k}},\psi^T_{\bm{-k}})$, where 
\begin{equation}
\psi^T_{\bm{k}} = (d_{z^2,\uparrow},d_{xy,\uparrow},d_{x^2-y^2,\uparrow}, d_{z^2,\downarrow},d_{xy,\downarrow},d_{x^2-y^2,\downarrow}).
\end{equation}
Here, $d_{\alpha,\sigma}(\bm{k})$ are annihilation operators for electrons in orbital $\alpha$ and spin $\sigma$ (we have suppressed the momentum index above). $\Delta$ in Eq.~\eqref{Hsc} is a $6 \cross 6$ matrix representing pairing between each possible combination of spin and orbital states. For simplicity, we assume that $\Delta$ is momentum-independent (i.e., the pairing potential is on-site in real space). This is justified as a first approximation since the Coulomb screening length is comparable to the lattice constant, and electron-phonon coupling is typically momentum-independent; hence, the interactions that give rise to pairing are most significant between pairs of states on the same site.

In a single-orbital system, the only momentum-independent pairing term allowed by symmetry is the trivial $s$-wave term $\Delta(\bm{k}) = i\sigma_y$, where $\sigma_i$ are the Pauli matrices in the spin degree of freedom. This is because fermion anticommutativity requires that
\begin{equation}
\label{DeltaIdentity}
\Delta(\bm{k}) = -\Delta^T(-\bm{k}),
\end{equation}
so a pairing matrix with no $k$-dependence and only a spin degree of freedom must be antisymmetric in spin - a spin-singlet. However, multi-orbital systems have an additional degree of freedom, so in this case, on-site spin-triplet pairing is possible, as long as this term is antisymmetric in the orbital degree of freedom. 

The most general momentum-independent pairing matrix allowed by the symmetry of our system has the form
\begin{equation} 
\label{Delta}
\Delta =\left[\begin{array}{cccccc}
0 & \Delta_4 & i\Delta_4 & \Delta_1 & 0 & 0\\
-\Delta_4 & 0 & 0 & 0 & \Delta_2 & i\Delta_3\\
-i\Delta_4 & 0 & 0 & 0 & -i\Delta_3 & \Delta_2\\
-\Delta_1 & 0 & 0 & 0 & \Delta_4 & -i\Delta_4\\
0 & -\Delta_2 & i\Delta_3 & -\Delta_4 & 0 & 0\\
0 & -i\Delta_3 & -\Delta_2 & i\Delta_4 & 0 & 0
\end{array}\right],
\end{equation} 
where $\Delta_{\gamma=1,\dots,4}$ are real parameters (though time-reversal symmetry breaking can allow for additional terms, including imaginary components of some of these parameters). 
$\Delta_1$ and $\Delta_2$ represent intra-orbital singlet pairing in the $d_{z^2}$ orbital and in the in-plane orbitals ($d_{x^2-y^2}$ and $d_{xy}$), respectively. $\Delta_3$ is an inter-orbital triplet which pairs the two in-plane orbitals. Our work focuses primarily on $\Delta_4$, another inter-orbital triplet term which notably pairs states of the same spin (it occupies the diagonal blocks of Eq.~\eqref{Delta}).

Of the four pairing terms, $\Delta_{1,2,3}$ are even under $M_z$, but $\Delta_4$ is odd. Thus, a state where both $\Delta_4$ and any of $\Delta_{1,2,3}$ are non-zero breaks $M_z$ symmetry. Such a state is only possible when the system is at the surface of a substrate, or in the top layer of a three-dimensional material such as 4Hb-TaS$_2$~\footnote{In the bulk of 4Hb-TaS$_2$, there is a mirror symmetry with respect to each 1H-TaS$_2$ layer, so this state is forbidden (unless $M_z$ is broken spontaneously)}. As we will see in Sec.~\ref{exp}, such breaking of $M_z$ is important in our explanation of the experimental results measured by Nayak \textit{et al} \cite{Beidenkopf2021}.

For now, we will not attempt to derive the values of the pairing terms from a microscopic model. Instead, we will examine the effect of setting various values for the pairing terms in order to characterize the unconventional SC phases that are possible in our system. We will return to the question of which pairing terms are favored by the system, and under which conditions, in Sec.~\ref{rg}.

\subsection{Nodal Phase}
\label{nodal}

\begin{figure}
     \subfloat[]{\includegraphics[width=0.45\textwidth]{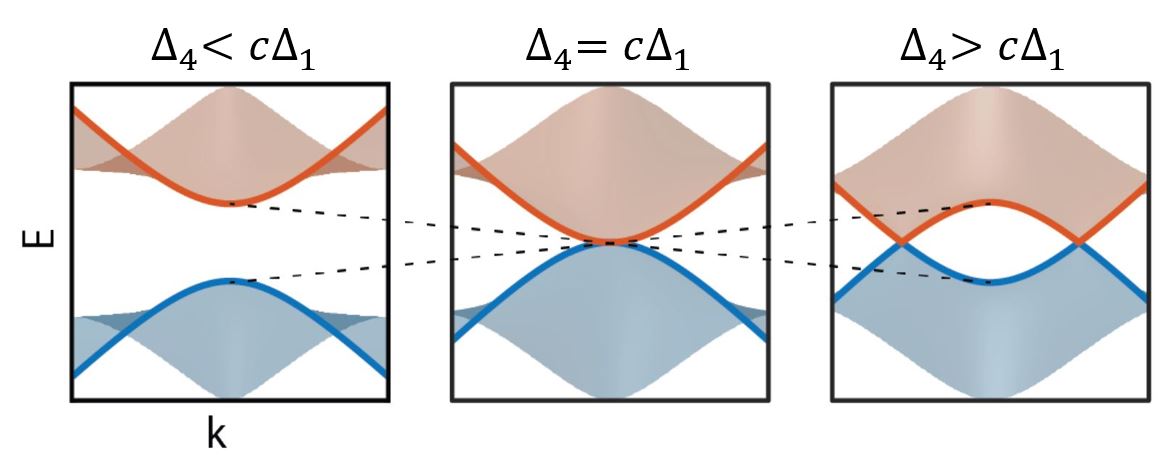}
        \label{nodes}}
        
     \subfloat[]{\includegraphics[width=0.22\textwidth]{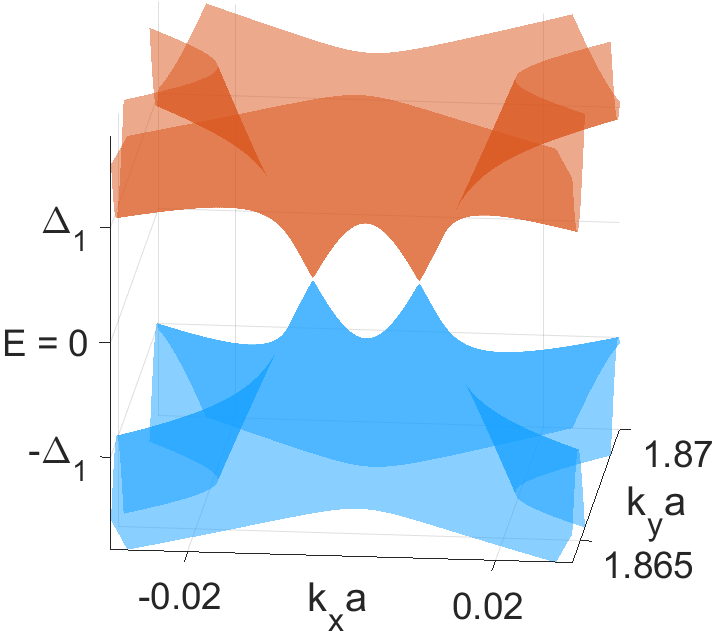}
        \label{dirac}}
     \subfloat[]{\includegraphics[width=0.23\textwidth]{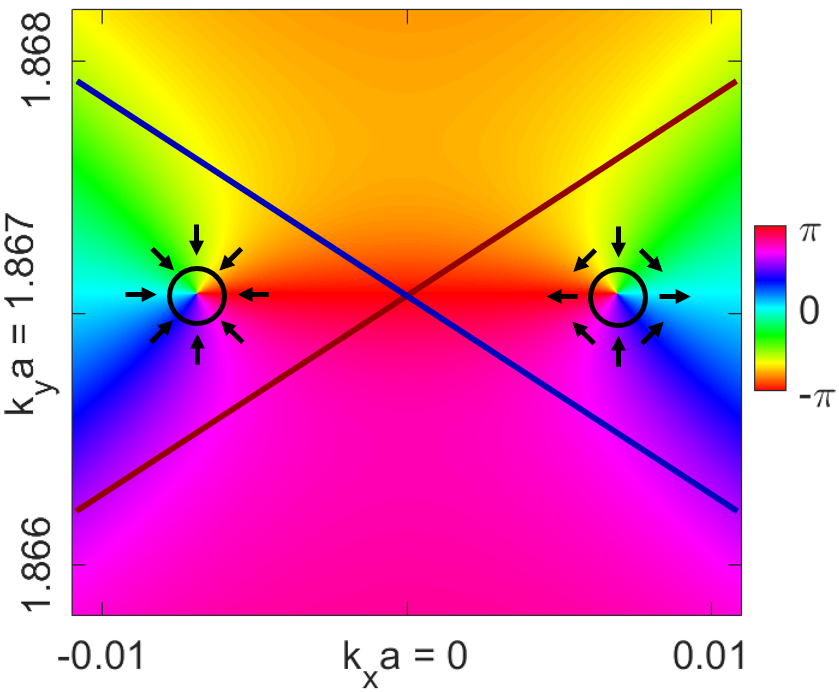}
        \label{winding}}
        
        \caption{(a) Schematic of the band structure at a band crossing for various values of $\Delta_4$. The constant $c \approx 3.32$ is dependent on the TaS$_2$ band structure parameters. (b) Modeled band structure near the band crossing in the positive $k_y$ direction from $\Gamma$. Here, $\Delta_4 = 1.5c\Delta_1$. (c) Color plot where color indicates the value of the characteristic angle $\theta_{\bm{k}}$ (defined in Appendix~\ref{theta}) near a band crossing (blue and red lines show the relevant bands). The positions of the two nodes nearest this crossing are circled. The characteristic angle has a winding of +1 (-1) around the left (right) node. Vectors $(\text{cos}\theta_{\bm{k}},\text{sin}\theta_{\bm{k}})$ are plotted near the nodes to illustrate their opposite windings.}
\end{figure}

As the singlet terms ($\Delta_1$ and $\Delta_2$) and the $\Delta_4$ term are the most relevant in explaining the experiment~\cite{Beidenkopf2021}, we set $\Delta_3$ to 0 in this analysis. Furthermore, for simplicity, we also set $\Delta_1=\Delta_2$ and neglect the Rashba SOC term for the time being (in our numerical analysis later in Sec.~\ref{rg}, we will relax these restrictions and allow all four terms to take on any value). For the case where $\Delta_4$ is 0, the system is a trivial $s$-wave SC, with a gap of size $\Delta_1$ over the entire Fermi surface. However, when $\Delta_4$ is finite, it suppresses this gap near the band crossings (circled in Fig.~\ref{fermi}) as shown schematically in Fig. \ref{nodes}. 
When $\Delta_4 = c \Delta_1$ (where $c \approx 3.32$ is a dimensionless constant determined by the band structure), the gap closes. For larger $\Delta_4$, a nodal phase is formed, with two nodal points near each band crossing (for a total of 12 nodes in the entire Brillouin zone). Near each node, the spectrum is linear, forming a Dirac dispersion, as plotted in Fig.~\ref{dirac}. The nodal points are protected by time reversal symmetry, which prevents the nodes from either shifting away from zero energy or opening a gap~\cite{Sato2006,Berg2008,Beri2010}.

Note that the nodes do not reside at high-symmetry points, and therefore their locations depend continuously on parameters. The momentum separation between the nodes is given by $\frac{2}{v_f}\sqrt{\Delta^2_4 - c^2\Delta^2_1}$, where $v_f$ is the Fermi velocity. This separation is much smaller than the Fermi momentum, which is of order $\frac{E_F}{v_F}$. Thus, in the absence of a significant Rashba SOC term, the nodes remain close to the midpoint of each side of the central hexagonal Fermi pocket. 
The effect of Rashba SOC will be considered in Sec.~\ref{tritops}.

\begin{figure}
     \subfloat[]{\includegraphics[width=0.178\textwidth]{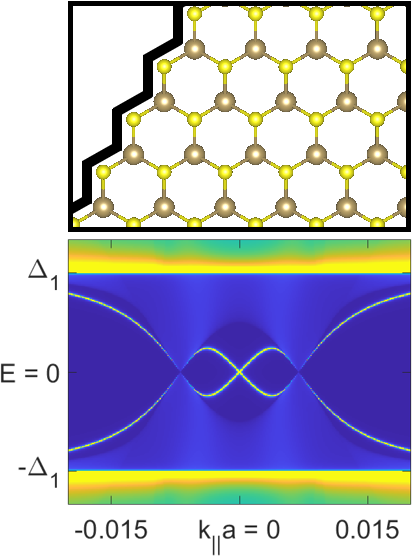}
        \label{zigzag}}
     \subfloat[]{\includegraphics[width=0.15\textwidth]{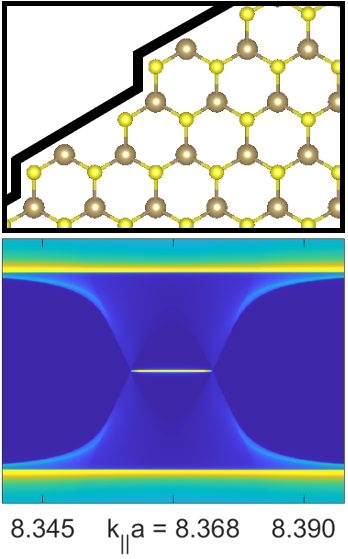}
        \label{slope1}}
     \subfloat[]{\includegraphics[width=0.15\textwidth]{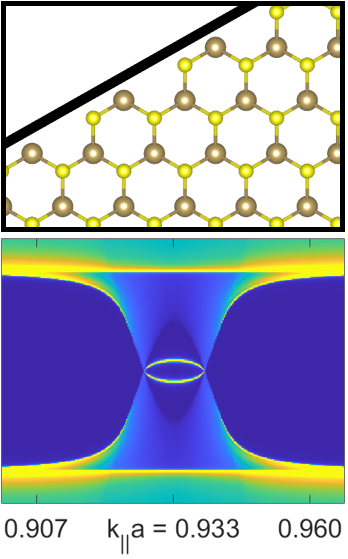}
        \label{armchair}}
        
     \subfloat[]{\includegraphics[width=0.255\textwidth]{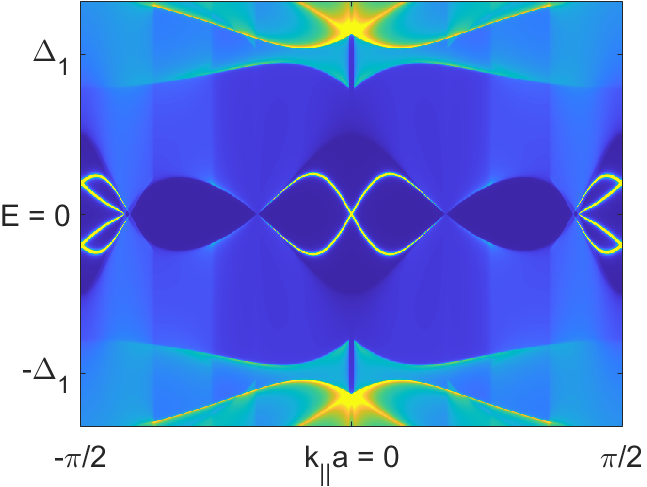}
        \label{alpha1}}
     \subfloat[]{\includegraphics[width=0.233\textwidth]{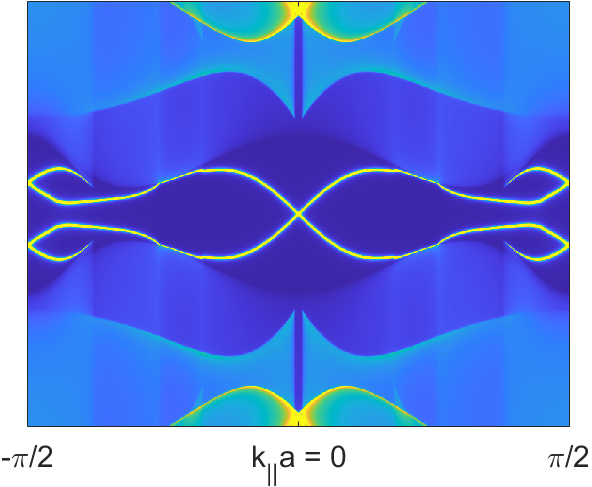}
        \label{alpha2}}
        
        \caption{Density of states (DOS) plots projected to the system edge, plotted as a function of momentum parallel to the edge $k_{||}$ (multiplied by the lattice constant $a$) and energy $E$. Due to projection to the edge, discrete edge modes have higher intensity; a lower-intensity bulk continuum is also visible.  (a) DOS projected to the zigzag edge (pictured above). Two edge modes can be seen crossing between the Dirac cones. (b) DOS projected to an edge direction which does not pass through any high-symmetry points other than $\Gamma$. This prevents the Dirac cones from overlapping in the projection, allowing an edge mode between a single pair of nodes to be viewed in isolation. Time-reversal and particle-hole symmetry pin the mode to zero energy. (c) DOS projected to the armchair edge. This edge has no contribution from edge modes at $E = 0$, demonstrating anisotropy between different edge directions. (d) A Rashba spin-orbit coupling (SOC) term of energy 10 meV is added, widening the distance between the cones, as shown in this DOS plot of the zigzag edge (parallel to $k_x$). Note the difference in scale between this and figure (a), which has no Rashba SOC. (e) The same plot as (d) but with the Rashba term increased to 20 meV. When SOC exceeds a critical value, the cones which had originated from the midpoints of the central hexagon's sides meet at the corners and merge, gapping out the full Brillouin zone. This causes the onset of the gapped TRITOPS phase.}
        \label{dos}
\end{figure}

When a pair of Dirac nodes form, each carries an opposite topological charge~\cite{Beri2010}, which we illustrate in Fig.~\ref{winding} as the winding of the angle $\theta_{\bm{k}}$ (defined in Eq.~\eqref{eq:thetak}) about the nodal points. The onset of nontrivial topology can also be seen in the appearance of Bogoliubov bound states in certain edge directions, which arise at the transition to the nodal phase. Such modes are depicted in Fig.~\ref{dos}, which displays the momentum-resolved, edge-projected density of states (DOS) $N(k,E)=-\frac{1}{\pi}\text{Im}(G_{00}(k,E))$. Here, $k$ is momentum along the edge and $G_{00}$ is the Green's function projected to the row of unit cells closest to the edge, derived using the method of Ref.~\cite{Sancho1985}.

Fig.~\ref{zigzag} shows the DOS projected to a zigzag edge of the system, while Fig.~\ref{armchair} shows an armchair edge. Only the former has zero-energy edge states, due to a crossing conserved by time-reversal symmetry. For an arbitrary edge direction in which the Dirac cones are not aligned (Fig.~\ref{slope1}), zero-energy modes appear in a range of momenta along the edge between each pair of opposite-charge nodes. These modes are pinned to zero energy by the combination of time reversal and particle-hole symmetries, which guarantees that every state at fixed $k$ and nonzero energy $E$ must have a partner with the same momentum at energy $-E$.

\subsection{Rashba SOC and the TRITOPS Phase}
\label{tritops}

In the previous analysis, we assumed that $M_z$-breaking at the substrate surface permitted the existence of a large $\Delta_4$ pairing term. However, we did not add any explicitly-$M_z$-breaking term to the Hamiltonian aside from $\Delta_4$. We now add such a term in the form of Rashba SOC, which can be viewed as an effective electric field term generated by the presence of charge on one side of the monolayer plane that is not matched on the other side.

This TR-invariant term has the form $H_{Rashba}(\bm{k}) = i\alpha\sum_{\bm{t}} (t_x\sigma_y - t_y\sigma_x)\text{exp}(i \bm{t} \cdot \bm{k} a)$, where $\alpha$ is a coefficient with units of energy, each $\bm{t}$ represents a lattice unit vector between nearest-neighbor sites, $a$ is the lattice constant, and $\sigma_i$ are Pauli matrices in the spin degree of freedom (we treat this term as orbital-independent, with an identity matrix in the orbital degree of freedom). This term is added to the 6-band bare Hamiltonian $H_0(\bm{k})$ described in Sec. \ref{model}.

The primary effect of Rashba SOC on the model is to enhance the separation between the Dirac cones; close to the band crossing, the separation increases by a factor of $\frac{\sqrt{\alpha^2+\Delta_1^2}}{\Delta_1}$. Since Rashba SOC is typically on the order of tens of meV, while $\Delta_1$ is less than 1 meV, this is a very large effect, and can result in the separation being of the same order as the width of the Brillouin zone (see Fig.~\ref{alpha1}).

As $\alpha$ increases, the cones move apart, starting at the centers of each side of the central hexagonal curves of the Fermi surface and moving toward the corners. For $\Delta_1 = 0.45$ meV and $\Delta_4= 1.5 c\Delta_1$, we find that when $\alpha$ exceeds $\alpha_C \approx 15$ meV, the cones merge at the corners, and the system becomes fully gapped. However, this merging does not return the system to a trivial phase; instead, the edge modes from the nodal phase remain, forming a class of topological superconductor known as a TRITOPS, or time-reversal-invariant topological superconductor.

This phase is characterized by a single pair of counter-propagating chiral Majorana edge modes which is protected by time-reversal symmetry. This is in constrast to the nodal phase, where the edge states are not topologically protected due to the absence of a gap, and hence can be disrupted by high-momentum impurity scattering that mixes the edge and the bulk. We demonstrate the nontrivial topology of the TRITOPS phase in the next section by calculating the topological invariant.


\subsection{Topological Invariant}
\label{topinv}

Since this system is 2-dimensional and in class DIII (with both particle-hole symmetry and time-reversal that squares to $-1$), we can define a topological invariant in $\mathbb{Z}_2$ when the system is fully gapped. In this section, we calculate the invariant to confirm that the state that arises when the nodes merge is indeed topological. A quantity related to this invariant is also meaningful in the nodal regime, elegantly relating the positions of the nodes to the topology of the system.

The $\mathbb{Z}_2$ invariant can be computed exactly in terms of the matrix $Q_{\bm{k}}$ defined in Appendix \ref{theta} \cite{Schnyder2008}, which is also used to calculate the characteristic angle $\theta_{\bm{k}}$ plotted in Fig. \ref{winding}. However, in the weak-pairing limit where $\Delta$ is orders of magnitude smaller than the energy differences between bands, the invariant can be expressed in a much simpler and more intuitive form.

\begin{figure}
     \subfloat[]{\includegraphics[width=0.22\textwidth]{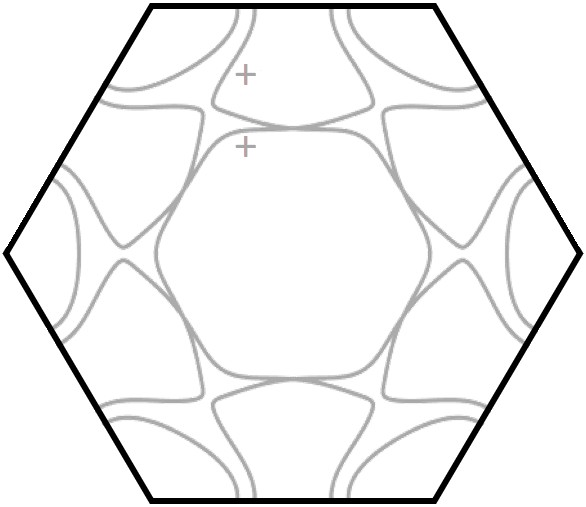}
        \label{fermi0}}
     \subfloat[]{\includegraphics[width=0.22\textwidth]{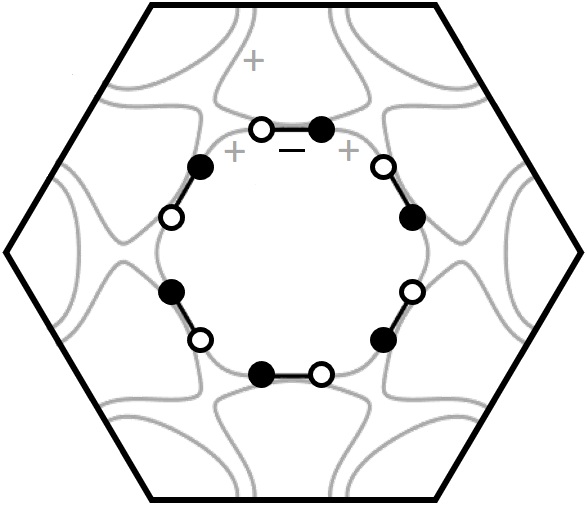}
        \label{fermi1}}

     \subfloat[]{\includegraphics[width=0.22\textwidth]{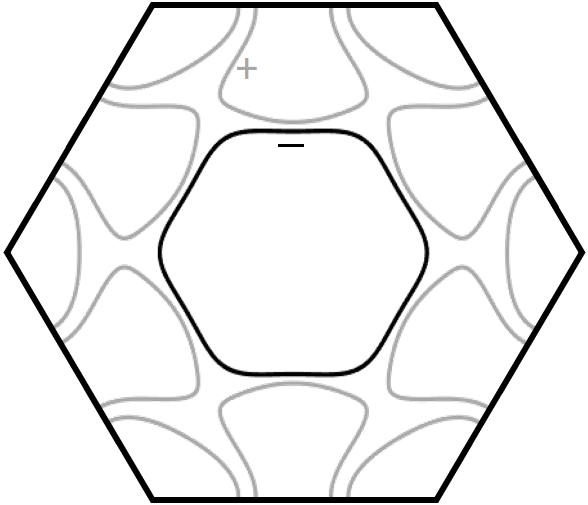}
        \label{fermi2}}
        
        \caption{Brillouin zone for (a) Rashba energy $\alpha = 0$, (b) $\alpha = \frac{1}{2}\alpha_C$, and (c) $\alpha = \alpha_C$, where, for the pairing energies used, the critical Rashba energy $\alpha_C = 15$ meV. Fermi lines colored gray (black) have positive (negative) effective pairing. The position of each node of winding +1 (-1) is indicated with a white (black) circle in (b).}
        \label{invariants}
\end{figure}

It has been shown \cite{SCZhang2010} that in this limit, the topological invariant for a TRITOPS is equal to the parity of the number of Fermi pockets with negative effective pairing, where the effective pairing for any band that crosses the Fermi energy is $\delta_{b,\bm{k}} = \bra{b,\bm{k}}\Delta\ket{b,\bm{k}}$. Here, $\ket{b,\bm{k}}$ is the $b$th eigenstate of $H_0(\bm{k})$ from Eq.~\eqref{Hsc}. Time-reversal symmetry guarantees that $\delta_{b,\bm{k}}$ can be chosen to be real. When the TRITOPS is fully gapped, all points $\bm{k}$ along a Fermi pocket must have the same sign of effective pairing, so this is a well-defined $\mathbb{Z}_2$ invariant.

In the nodal regime, the invariant is not well-defined. However, as the nodes travel along the Fermi lines, $\delta_{b,\bm{k}}$ acquires a sign between them, as shown in Fig.~\ref{invariants}. Thus, it demonstrates how, between the creation and merging of the nodes, the nodal phase acts as an intermediate stage between the trivial and topological SC phases.

The presence of an intermediate nodal phase is highly generic \cite{DasSarma2021}. In any system that transitions from a trivial superconductor to a TRITOPS in the weak pairing limit, the effective pairing of one or more Fermi pockets must change sign. Unless a symmetry constrains this sign change to occur simultaneously at all points along the pocket (such as in artificial models which are rotationally symmetric due to considering only momenta close to $\Gamma$), we expect nodes to appear to mediate the sign change.

\section{Relation to Experiment}
\label{exp}

This work is motivated by an experiment by Nayak \textit{et al.} \cite{Beidenkopf2021} in which an increased DOS was observed on step edges on the surface of superconducting 4Hb-TaS$_2$ via STM measurements.

The 4Hb lattice consists of alternating layers of 1H-TaS$_2$, which is conducting, and 1T-TaS$_2$, which is insulating but hosts a significant charge density wave (CDW). Due to weak coupling between the layers, superconductivity can be seen as confined primarily to the 1H layers. 

Sub-gap edge DOS was observed both at downward steps (where a terminating 1H layer ends, revealing the 1T layer below it) and at upward steps (where a terminating 1T layer begins, covering the underlying 1H layer). 
The former observation may be explained by assuming that the superconducting 1H layer has a sub-gap edge state at its boundary with vacuum. However, this cannot explain the existence of an edge state at upward steps, since the 1H layer does not terminate there, but rather continues beneath the 1T layer.

To account for this, we proposed~\cite{Beidenkopf2021} that the edge state at the upward step arises due to the breaking of $M_z$ symmetry in the 1H layer on the surface. This symmetry breaking is stronger when the 1H layer is the topmost layer. Therefore, the region of the 1H layer which is exposed can be in a different phase than the region that has a 1T layer above it, and as a result, an edge state appears on the boundary. Our model accounts for this since, as we have demonstrated, $M_z$ breaking can drive the system from the trivial to the nodal, and then to the gapped topological phase. Furthermore, our model predicts sub-gap modes to appear between these phases.


Another relevant experimental finding is anisotropy of the localization length, where edge modes along differently-oriented step edges decay into the bulk over different length scales. Our model predicts different projected DOS structure along different edge directions (as can be seen by comparing Figs. \ref{zigzag} and \ref{armchair}), which lines up qualitatively with the experimental result.

We note, however, that some features of the experimental data are not reproduced by our model. First, the DOS away from the step edges does not go to zero at the Fermi level, as it would for either a fully-gapped or a nodal superconductor. Second, the localization length of the edge modes obtained in the experiment are about an order of magnitude shorter than those predicted theoretically~\cite{Beidenkopf2021}. Disorder (either magnetic or non-magnetic~\cite{Dentelski2021}), which is not accounted for in our theory, may play a role in both of these effects. Specifically, disorder tends to shorten the coherence length in superconductors and to induce a finite DOS at zero energy, so it may be sufficient to explain the discrepancies between our model and observations.

\section{Nodal and topological phases from microscopic interacting model}
\label{rg}

\subsection{Model}
We now turn to the crucial question of which microscopic interactions can give rise to multi-orbital topological and nodal pairing in TMD systems. We begin with the action
\begin{equation}
    S = S_0 + S_U.
\end{equation}
Here, $S_0$ is the non-interacting part of the action,
\begin{equation}
S_0 = \sum_{\omega_n,\bm{k}}  \psi_{\bm{k},\omega_n,i}^\dagger\big[i\omega_n\delta_{ij} - [H_0]_{ij}(\bm{k})\big]\psi_{\bm{k},\omega_n,j},  
\end{equation}
where $\omega_n$ are fermionic Matsubara frequencies and $H_0$ is the non-interacting Hamiltonian in Eq.~\eqref{Hsc}. The indices $i,j$ are summed from 1 to 6, representing both spin and orbital degrees of freedom. Summation over repeated indices is implied.

The interacting part of the action is given by
\begin{equation}
\begin{aligned}
\label{SU}
    S_U &= S_C + S_J + S_{ph} \\
    S_C &= \frac{C}{2}\int d\tau \sum_{\bm{R}} \bigg[ \sum_{m,s} n_{m,s}(\tau,\bm{R}) \bigg]^2 \\
    S_J &= -\frac{J}{2}\int d\tau \sum_{\bm{R}} \sum_{m \neq m'} \vec S_m (\tau,\bm{R}) \cdot \vec S_{m'} (\tau,\bm{R}) \\
    S_{ph} &= -\frac{1}{2}\int d\tau d\tau' g(\tau - \tau') \ \cross \\
    &\quad \sum_{\bm{R}} \sum_{m,m',s,s'} n_{m,s}(\tau,\bm{R}) n_{m',s'}(\tau',\bm{R}),
\end{aligned}
\end{equation}
where $m$ and $m'$ are orbital indices, $s$ and $s'$ are spin indices, $\tau$ and $\bm{R}$ are imaginary time and position, respectively, and
\begin{equation}
\begin{aligned}
    n_{m,s} (\tau,\bm{R}) &= \psi_{\bm{R},\tau,m,s}^\dagger \psi_{\bm{R},\tau,m,s} \\
    \vec S_m (\tau,\bm{R}) &= \psi_{\bm{R},\tau,m,s}^\dagger \vec\sigma_{s,s'} \psi_{\bm{R},\tau,m,s'}
\end{aligned}
\end{equation}
for $\vec\sigma$ the vector of Pauli matrices $(\sigma_x, \sigma_y, \sigma_z)$.

The above definition of the interacting terms uses the angular momentum basis $m = \{d_0, d_{+2},d_{-2}\}$, rather than the Cartesian basis $m = \{d_{z^2}, d_{xy},d_{x^2-y^2}\}$, for its orbital degree of freedom. Here, $d_0 = d_{z^2}$ and $d_{\pm 2} = d_{x^2-y^2} \pm i d_{xy}$. We then apply a basis transformation to the Cartesian basis (see Appendix~\ref{basis}) to make the terms compatible with our previously-computed Hamiltonian.

The three interaction terms we consider are $S_C$, a local density-density coulomb repulsion term, $S_J$, an exchange term that enforces Hund's rule, and $S_{ph}$, a phonon-mediated attraction term.

$S_C$ is spin and orbital-independent, with a coupling strength given by $C$. $S_J$, proportional to the positive coupling strength $J$, favors spins aligning when they are in different orbitals. Neither of these terms are frequency-dependent.

For $S_{ph}$, we assume a constant attraction below the Debye frequency and no effect for higher frequencies. That is,
\begin{equation}
    g(\tau) = T \sum_{\Omega_n} e^{-i\Omega_n\tau} \tilde{g}(\Omega_n),
\end{equation}
where
\begin{equation}
    \tilde{g}(\Omega_n) =
    \begin{cases} 
      g, & |\Omega_n| \leq \omega_D \\
      0, & |\Omega_n| > \omega_D.
   \end{cases}
\end{equation}
Here, $g$ is the electron-phonon coupling strength, $T$ is temperature, $\Omega_n$ are bosonic Matsubara frequencies, and $\omega_D$ is the Debye frequency.

In order to use the action to compute $\Delta$, we combine these three terms into a single, general interaction tensor in the Cartesian basis, of the form
\begin{equation}
\begin{aligned}
\label{Utensor}
S_U = \sum_{\omega_n,\omega'_n,\Omega_n} \sum_{\bm{k},\bm{k}',\bm{q}} &\psi_{\bm{k},\omega_n,i}^\dagger \psi_{\bm{k}',\omega'_n,j}^\dagger U_{ijkl}(\Omega_n) \ \cross \\
&\psi_{\bm{k}'+\bm{q},\omega'_n+\Omega_n,k} \psi_{\bm{k}-\bm{q},\omega_n-\Omega_n,l}.
\end{aligned}
\end{equation}

In the next section, we use an RG procedure to integrate out the high-frequency dependence of $U_{ijkl}(\Omega_n)$, allowing us to work in the regime below $\omega_D$, where it is constant. This enables us to solve for $\Delta$ self-consistently via a generalization of the gap equation:
\begin{equation}
\label{gapeq}
\Delta_{ij} = -U_{ijkl} F_{kl},
\end{equation}
where $F$ is the anomalous Green's function. It has the form
\begin{equation}
F =T\sum_{\omega_n}\int \frac{d^2k}{(2\pi)^2} \big[(i\omega_n - H_{SC}(\bm{k}))^{-1}\big]_{12}.
\end{equation}
The subscript $12$ refers to the top-right particle-hole block, making $F$ a $6 \cross 6$ matrix for this system. The anomalous Green's function can be computed numerically given the BdG Hamiltonian $H_{SC}(\bm{k})$.

If we know the interaction tensor $U$, we can thus solve for $\Delta$ iteratively by assuming an initial matrix $\Delta$, constructing $H_{SC}$ from this matrix, computing $F$, and then solving Eq~\eqref{gapeq} to get an updated $\Delta$; this is repeated until convergence.

\subsection{Interaction Tensor RG}
\label{itrg}

In this section, we construct an effective interaction tensor at the Debye energy scale. This generalizes the formalism of Anderson, Morel, and Tolmachev \cite{Anderson1962,Tolmachev1961} to a multi-orbital system.

Since $\Delta$ and $F$ in Eq.~\eqref{gapeq} transform in the same way under the symmetries of the system, we can always express $F$ as a linear combination of the four basis matrices which compose $\Delta$. We designate these matrices $X_{\gamma}$ for $\gamma=1,\dots,4$, where each matrix corresponds to the coefficient $\Delta_{\gamma}$. By definition, $\Delta = \Delta_{\gamma} X_{\gamma}$, and similarly, we can write $F = F_{\gamma} X_{\gamma}$.

This allows us to simplify our calculation by expressing $U$ as a $4\cross4$ matrix in the basis $\{X_1, X_2, X_3, X_4\}$, rather than as a $6^4$-element tensor. The transformation is
\begin{equation}
\label{transform4x4}
\hat{U}_{\gamma\delta} = (X^{*}_{\gamma})_{ij} U_{ijkl} (X_{\delta})_{kl}.
\end{equation}

In this basis, $\hat{U}(\Lambda_0)$, the interaction tensor at the UV cutoff energy scale $\Lambda_0$, turns out to simply be
\begin{eqnarray}
\hat{U}(\Lambda_0) &=& \hat{U}^C + \hat{U}^J \nonumber \\
&=& \left[\begin{array}{cccc}
C & 0 & 0 & 0\\
0 & C + 3J & 0 & 0\\
0 & 0 & C - J & 0\\
0 & 0 & 0 & C - J
\end{array}\right].
\end{eqnarray}

We see that while Coulomb repulsion affects all pairing terms equally, exchange favors the triplet terms $\Delta_3$ and $\Delta_4$ while strongly suppressing the in-plane singlet, $\Delta_2$.

As we derive in Appendix~\ref{derivation}, $\hat{U}$ at any energy scale $\Lambda$ can be computed from $\hat{U}(\Lambda_0)$ by integrating the generalized RG equation
\begin{equation} \label{rgeq}
\hat{U}^{-1}(\Lambda-d\Lambda) - \hat{U}^{-1}(\Lambda) = \frac{d\Lambda}{\Lambda}\hat{\nu}_C(\Lambda),
\end{equation}
where $\hat{\nu}_C(\Lambda)$ is a tensor reflecting the density of states in the Cooper channel as a function of energy scale. In the single-band case with only one pairing channel, $\hat{\nu}_C(\Lambda)$ reduces exactly to the density of states at energy $\Lambda$; in systems with multiple channels, the eigenvalues of $\hat{\nu}_C$ indicate the strengths of effective pairing channels in the basis where they are uncoupled. $\hat{\nu}_C$ can be computed from the Hamiltonian as
\begin{equation}
\begin{aligned}
\label{nueq}
(\hat{\nu}_C)_{\gamma \delta}(\Lambda) = \sum_b \int \frac{d^2k}{(2\pi)^2} \Big[&\delta(|E_b(\bm{k})|-\Lambda) \ \cross \\
&(\tilde{X}^{\dagger}_{\gamma})_{bb}(\bm{k}) (\tilde{X}_{\delta})_{bb}(\bm{k})\Big],
\end{aligned}
\end{equation}
where
\begin{equation}
\tilde{X}_{\gamma}(\bm{k}) = \mathcal{U}^T(-\bm{k}) X_{\gamma} \mathcal{U}(\bm{k})
\end{equation}
is the transformation of $X_{\gamma}$ into the eigenbasis of $H_0(\bm{k})$. The index $b$ enumerates the 6 bands of $H_0(\bm{k})$ with energies $E_b(\bm{k})$, while the eigenvectors at momentum $\bm{k}$ form the columns of $\mathcal{U}(\bm{k})$.

\begin{figure}
     {\includegraphics[width=0.45\textwidth]{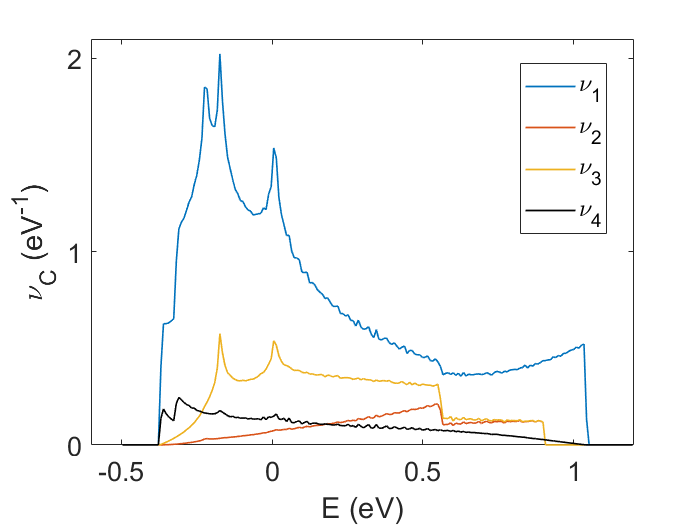}}
        
        \caption{Eigenvalues $\nu_{\gamma}$ of the pairing density $\hat{\nu}_C$ as a function of energy scale. $\nu_1$ and $\nu_2$ are intra-orbital $s$-wave pairing channels that are linear combinations of the $\Delta_1$ and $\Delta_2$ channels; they can be considered, respectively, the sum and difference of the pairing strengths of the $d_{z^2}$ orbital and the in-plane orbitals ($d_{xy}$ and $d_{x^2-y^2}$). $\nu_3$ and $\nu_4$ can be straightforwardly interpreted as the the strengths of the $\Delta_3$ and $\Delta_4$ pairing channels, respectively, as these channels have very little coupling to each other or to the other two terms. In the above plot, sharp peaks or drops in density relate to band edges.}
        \label{sus}
\end{figure}

The four eigenvalues of $\hat{\nu}_C$ are plotted in Fig.~\ref{sus} as a function of energy (for this plot, we use a physically-reasonable Rasbha term of $\alpha = 10$ meV). To good approximation, $\hat{\nu}_C$ is diagonal aside from strong coupling between the $\Delta_1$ and $\Delta_2$ channels, so the third and fourth eigenvalues $\nu_3$ and $\nu_4$ closely reflect the strength of pairing in the $\Delta_3$ and $\Delta_4$ channels, respectively.

The $s$-wave channels, however, hybridize due to the coupling between them. Rather than referring to the $\Delta_1$ and $\Delta_2$ channels, $\nu_1$ and $\nu_2$ instead approximately indicate the sum and difference of these channels, respectively. Thus, the comparison most relevant to this work is between $\nu_1$, the total $s$-wave intra-orbital pairing, and $\nu_4$, the inter-orbital same-spin triplet pairing.

As can be seen in the figure, $\nu_1 > \nu_4$ at every relevant energy. This is due to the fact that $\Delta_4$ pairs identical spins at opposite momenta, while Ising SOC restricts spins to be opposite at all points except the band crossings; for this reason, the number of states that contribute to this type of pairing is fundamentally limited compared with opposite-spin intra-orbital pairing channels. However, this can be overcome by sufficiently strong interactions that favor triplet pairing, as we will show in Sec.~\ref{results}.

By integrating Eq.~\eqref{rgeq} with bounds at $\omega_D$ and $\Lambda_0$, we solve for $\hat{U}(\omega_D)$, the interaction tensor, at the relevant energy scale for phonon-mediated SC, in terms of the bare interaction $\hat{U}(\Lambda_0)$:
\begin{equation}
\hat{U}(\omega_D) = \Bigg[\hat{U}^{-1}(\Lambda_0) + \displaystyle\int\limits_{|\Lambda| > \omega_D}^{\Lambda_0} \frac{d\Lambda}{\Lambda}\hat{\nu}_C(\Lambda)\Bigg]^{-1}.
\end{equation}

For ease of interpretation, we will approximate $\hat{\nu}_C$ as fully diagonal in the discussion that follows. In this limit, the 4 pairing channels are fully decoupled, and the above equation reduces to 4 independent relations
\begin{equation}
\label{Ui}
U_{\gamma} \equiv \hat{U}(\omega_D)_{\gamma\gamma} = \Big[\hat{U}^{-1}_{\gamma\gamma}(\Lambda_0) + \Bar{\nu}_{\gamma}\text{log}\frac{\Lambda_0}{\omega_D}\Big]^{-1},
\end{equation}
where doubled indices denote a diagonal element rather than a sum, and
\begin{equation}
\label{nui}
\Bar{\nu}_\gamma\text{log}\frac{\Lambda_0}{\omega_D} \equiv \displaystyle\int\limits_{|\Lambda| > \omega_D}^{\Lambda_0} \frac{d\Lambda}{\Lambda}\nu_{\gamma}(\Lambda).
\end{equation}

Since we are now treating each pairing as independent, this at last leads to a simple form for the magnitude of each pairing channel $\Delta_\gamma$:
\begin{equation}
\label{Di}
\Delta_\gamma = 2\omega_D \text{exp}\bigg[\frac{1}{(U_\gamma - g)\bar{\nu}_\gamma}\bigg].
\end{equation}
Eq.~\eqref{Di} is the solution to the BCS gap equation assuming decoupled pairing channels. Indeed, substituting $U_{ijkl} = U_{ijkl}(\omega_D) - g\delta_{ik}\delta_{jl}$ into Eq.~\eqref{gapeq} and solving self-consistently yields close agreement for reasonable values of $C$, $J$, and $g$.

\subsection{Results}
\label{results}

The relative strengths of the different pairing channels determine the presence or absence of multi-orbital topological superconductivity. As we saw in the previous section, these strengths, in turn, are determined by two competing quantities: $\bar{\nu}_{\gamma}$, which characterizes the abundance of states in the band structure which can form pairs of the corresponding channel $\Delta_{\gamma}$, and $U_{\gamma}$, the repulsive interaction which must be overcome for pairing to occur in that channel. While the former strongly favors a combination of the $s$-wave pair potentials $\Delta_1$ and $\Delta_2$ over $\Delta_4$, the latter can suppress $s$-wave pairing more than $\Delta_4$ due to Hund's rule interactions.

As we will now show, within our simplified model, the nodal phase occurs only for a narrow range of parameters and with very low critical temperatures. To see this, consider the limit $J \approx C$ and $C\Bar{\nu}_1\text{log}\frac{\Lambda_0}{\omega_D} \gg 1$. Here, $U_4$ approaches 0 while $U_1$ approaches $[\Bar{\nu}_1\text{log}\frac{\Lambda_0}{\omega_D}]^{-1}$. For $\Lambda_0$ of the order of the band width ($\sim$~1 eV) and $\omega_D \approx 0.03$~ eV, this is approximately $0.3 \, \Bar{\nu}^{-1}_\gamma$.

Then, substituting Eq.~\eqref{Di} into the criterion for a nodal TSC, $\Delta_4 > c\Delta_1$ (see Sec.~\ref{nodal}), the condition becomes
\begin{equation}
\begin{aligned}
\label{ineq}
2\omega_D \text{exp}\bigg[-\frac{1}{g\Bar{\nu}_4}\bigg] &> 2c\omega_D \text{exp}\bigg[\frac{1}{0.3 - g\Bar{\nu}_1}\bigg].
\end{aligned}
\end{equation}
We can ignore the order-1 constant $c$ since its effect is small compared to the values in the exponents. Since $\Delta_1$ must be finite for the system to have a superconducting gap, we also require that $0.3 - g\Bar{\nu}_1 < 0$. The resulting inequality is
\begin{equation}
\label{gs}
\frac{0.3}{\Bar{\nu}_1} < g < \frac{0.3}{\Bar{\nu}_1 - \Bar{\nu}_4}.
\end{equation}
Since $\Bar{\nu}_4 \ll \Bar{\nu}_1$, we approximate this range for clarity as
\begin{equation}
\label{gs1}
\frac{0.3}{\Bar{\nu}_1} < g < \frac{0.3}{\Bar{\nu}_1} \bigg(1 + \frac{\Bar{\nu}_4}{\Bar{\nu}_1}\bigg).
\end{equation}

Not only is this a very fine-tuned window of possible interaction energies, but the resulting pair potentials $\Delta_1$ and $\Delta_4$ (substituting $g$ from Eq.~\eqref{gs1} into Eq.~\eqref{ineq}) are exponentially small, of order $\omega_D \, \text{exp}[-\frac{\Bar{\nu}_1}{\Bar{\nu}_4}]$. In this analytic estimate, off-diagonal coupling terms were neglected in the matrix $\hat{\nu}_C$. Including these terms and solving for $T_C$ numerically yields a result even smaller than this estimate.

To achieve the nodal phase at experimentally-attainable critical temperatures, we adjust the model to allow for electron-phonon coupling which is not uniform across all pairing channels. In the simplest such case, we keep $g$ constant in all channels other than the $\Delta_4$ pairing channel, where we instead use a larger phonon-mediated attraction term $g_4$.

\begin{figure}
     {\includegraphics[width=0.45\textwidth]{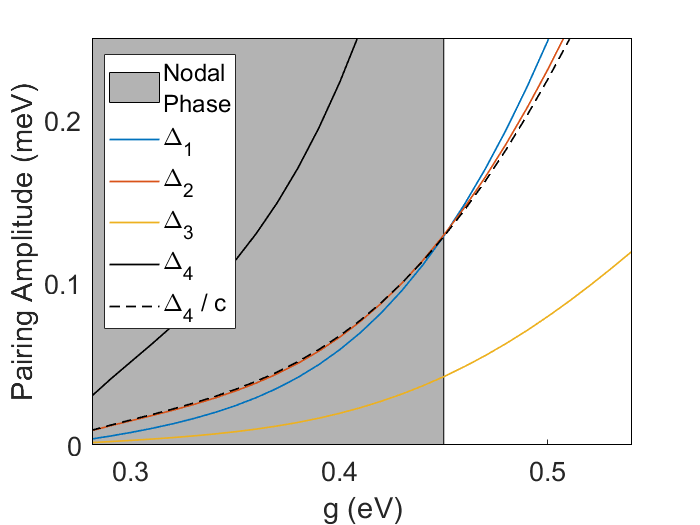}}
        
        \caption{Magnitudes of the 4 parameters of $\Delta$, computed by iterating the multi-orbital gap equation, where the Coulomb repulsion $C$ and Hund's rule interaction $J$ have energies of 0.5 and 0.4 eV, respectively. We also plot $\Delta_4 / c$ to show that the nodal SC phase (shaded region) transitions to a trivial SC phase (unshaded) when $\Delta_4 = c\Delta_1$ ($\Delta_1$ and $\Delta_2$ crossing near this point is unrelated and depends on the values of $C$ and $J$). In order to achieve the nodal phase for physically-reasonable values of $\Delta_1$ and $\Delta_4$ (order $\sim$0.1 meV), we take the electron-phonon coupling to be $5g$ in the $\Delta_4$ channel, while in all other channels it is $g$. If the coupling energy is instead taken to be identical across all channels, the crossing occurs at energies much smaller than are measurable.}
        \label{deltavsg}
\end{figure}

Fig.~\ref{deltavsg} illustrates the case $g_4 = 5g$ for $g$, $C$, and $J$ of the order of the bandwidth (the former is the domain of the plot, while the latter two values are 0.5 and 0.4 eV, respectively). The factor of 5 yields a reasonable order of magnitude ($\sim 10^{-4}$~ eV) for $\Delta_4$ and $\Delta_1$ in the regime where the condition for the nodal phase, $\Delta_4 > c\Delta_1$, is satisfied.

Reducing this factor exponentially suppresses the pairing amplitudes, and thus the critical temperature of the superconductor. For this reason, our model suggests that phonons must couple to the different pairing channels with different strengths in order to yield the nodal phase at observable temperatures.

\section{Conclusion}
\label{conc}

We have demonstrated that nodal and gapped topological TR-invariant superconductivity can occur in 1H-TMD layers. As an inter-orbital, $M_z$-breaking pairing term increases, we showed that a transition is induced, from the trivial to the nodal phase and finally to the gapped TSC, with each transition accompanied by the creation or annihilation of nodal points. Our theory was then related to a recent experiment in a layered TaS$_2$ system.

Our microscopic model suggested that, for a sufficiently large inter-orbital pair potential to arise (which is necessary to stabilize the nodal phase), we not only require strong Hund's rule interactions, but also a larger phonon-mediated attraction in the inter-orbital channel compared with the intra-orbital channel. This could result from, for instance, an electron-phonon coupling which is highly anisotropic in momentum. Since intra-orbital pairing has no momentum dependence, while the inter-orbital term is localized to band crossings (which occur along high-symmetry lines in momentum space), the electron-phonon coupling could then indeed be strongly channel-dependent. A more detailed exploration of phononic interactions in this model merits future study.

\section*{Acknowledgments}
\label{acknowledgments}

We acknowledge Haim Beidenkopf and Binghai Yan for useful discussions. We also thank David M\"ockli for his help in constructing the tight-binding model, and K.T. Law and Wenyu He for the use of their code to fit the band structure. G.M. and Y.O. were supported by the European Union's Horizon 2020 research and innovation programme (Grant Agreement LEGOTOP No. 788715), the DFG (CRC/Transregio 183, EI 519/7-1), ISF Quantum Science and Technology (2074/19), and the BSF and NSF grant (2018643). E.B. was supported by the European Research Council (ERC) under grant HQMAT (Grant Agreement No. 817799) the Minerva foundation, and a research grant from Irving and Cherna Moskowitz.

\appendix
\section{Tight-Binding Model}
\label{tb}

The tight-binding model used in this work consists of a hopping matrix for each of the 19 possible lattice vectors up to third-nearest-neighbor interactions. Each hopping matrix is a $6\cross6$ matrix in the space of the 2 spins ($\uparrow$ and $\downarrow$) and 3 orbitals ($d_{z^2}$, $d_{xy}$, and $d_{x^2-y^2}$) which are most relevant to the bands at the Fermi energy.

\begin{figure}
     {\includegraphics[width=0.45\textwidth]{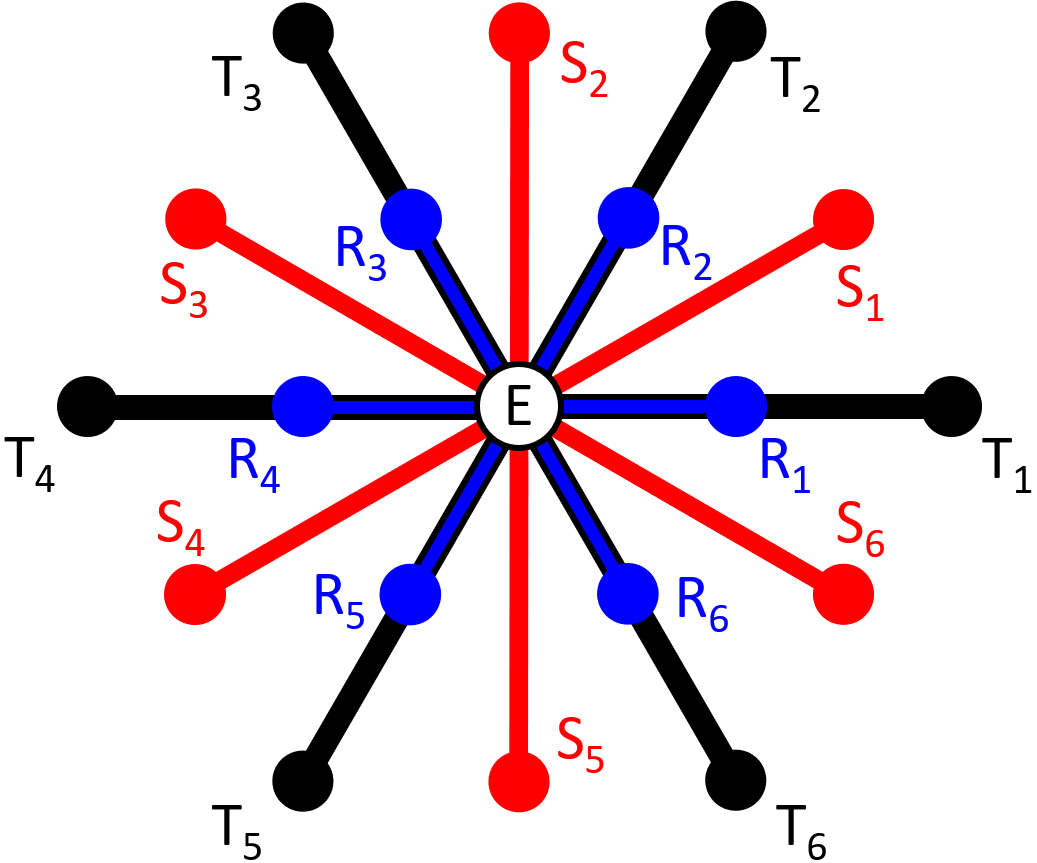}}
        
        \caption{Diagram of triangular lattice vectors up to third-nearest neighbor. We designate the on-site matrix $E$. The hopping matrix to each lattice vector $\bm{R}_i$, $\bm{S}_i$, or $\bm{T}_i$ is called $R_i$, $S_i$, or $T_i$, respectively.}
        \label{latticevectors}
\end{figure}

The resulting Hamiltonian has the form
\begin{eqnarray}
\label{Hk}
H_0(\bm{k}) &=& E + \sum_i^6 R_i e^{i \bm{R}_i \cdot \bm{k}} \nonumber \\
&+& \sum_i^6 S_i e^{i \bm{S}_i \cdot \bm{k}} + \sum_i^6 T_i e^{i \bm{T}_i \cdot \bm{k}},
\end{eqnarray}
where $E$ is the on-site hopping matrix, while the 6 $R_i$, $S_i$, and $T_i$ matrices are, respectively, the nearest-neighbor, next-nearest-neighbor, and third-nearest-neighbor hopping matrices, each corresponding to a lattice vector $\bm{R}_i$, $\bm{S}_i$, or $\bm{T}_i$. The lattice vectors in our naming convention are illustrated in Fig.~\ref{latticevectors}.

The on-site matrix $E$ is given by
\begin{eqnarray}
\label{E}
E &=& \sigma_0 \otimes \left[\begin{array}{ccc}
\epsilon_0 - \mu_0 & 0 & 0 \\
0 & \epsilon_1 - \mu_0 & 0 \\
0 & 0 & \epsilon_2 - \mu_0 \\
\end{array}\right] \nonumber \\
&+& \sigma_z \otimes \left[\begin{array}{ccc}
0 & 0 & 0 \\
0 & 0 & i\lambda_{SO} \\
0 & -i\lambda_{SO} & 0 \\
\end{array}\right],
\end{eqnarray}
where $\epsilon_0$, $\epsilon_1$, and $\epsilon_2$ are on-site energy terms, $\mu_0$ is a global chemical potential term, and $\lambda_{SO}$ is the amplitude of the Ising spin-orbit term (discussed in Sec.~\ref{model}). $C_3$ symmetry constrains $\epsilon_1$ and $\epsilon_2$ to have the same value.

The matrices $\sigma_0$ and $\sigma_z$ are the identity and the third Pauli matrix, respectively, in the spin degree of freedom, while the remaining $3\cross3$ matrices are in the orbital degree of freedom. All subsequent matrices in this section are $6\cross6$ and are written as a tensor product of a 2-dimensional spin matrix and a 3-dimensional orbital matrix.

The hopping matrices are restricted to contain only real parameters which satisfy the symmetries described in Sec.~\ref{model}, and it is assumed that no spin-orbit interaction is significant in the isolated monolayer other than the aforementioned Ising SOC. However, aside from these restrictions, the forms listed below are the most general-possible hopping matrices for a 1H-TMD.

Three of the hopping matrices are given by
\begin{equation}
\begin{aligned}
\label{R1S1T1}
R_1 &= \sigma_0 \otimes \left[\begin{array}{ccc}
t_0 & -t_1 & t_2 \\
t_1 & t_{11} & -t_{12} \\
t_2 & t_{12} & t_{22} \\
\end{array}\right] \\
S_1 &= \sigma_0 \otimes \left[\begin{array}{ccc}
r_0 & r_2 & -\frac{1}{\sqrt{3}}r_2 \\
r_1 & r_{11} & r_{12} \\
-\frac{1}{\sqrt{3}}r_1 & r_{12} & (r_{11}+\frac{2}{\sqrt{3}}r_{12}) \\
\end{array}\right] \\
T_1 &= \sigma_0 \otimes \left[\begin{array}{ccc}
u_0 & -u_1 & u_2 \\
u_1 & u_{11} & -u_{12} \\
u_2 & u_{12} & u_{22} \\
\end{array}\right].
\end{aligned}
\end{equation}

To simplify the definitions of the other 15 hopping matrices, we first define the $C_3$ matrix operator
\begin{equation}
C_3=\left[\begin{array}{cc}
e^{-i\frac{\pi}{3}} & 0\\
0 & e^{i\frac{\pi}{3}}
\end{array}\right] \otimes
\left[\begin{array}{ccc}
1 & 0 & 0 \\
0 & -\frac{1}{2} & \frac{\sqrt{3}}{2} \\
0 & -\frac{\sqrt{3}}{2} & -\frac{1}{2} \\
\end{array}\right].
\end{equation}
Note that $C_3$ both rotates spins and maps the in-plane orbitals $d_{xy}$ and $d_{x^2-y^2}$ onto each other. However, for the application in this section, the spin part always cancels to identity, so the manipulations that follow can be carried out exclusively in the $3\cross3$ orbital basis.

Using $C_3$, we can write all remaining hopping matrices as simple transformations of the previously-defined single matrices of each set:
\begin{equation}
\begin{aligned}
\label{C3relations}
R_2 &= C_3^\dagger R_1^\dagger C_3  & S_2 &= C_3^\dagger S_1^\dagger C_3 & T_2 &= C_3^\dagger T_1^\dagger C_3 \\
R_3 &= C_3 R_1 C_3^\dagger & S_3 &= C_3 S_1 C_3^\dagger & T_3 &= C_3 T_1 C_3^\dagger \\
R_4 &= R_1^\dagger & S_4 &= S_1^\dagger & T_4 &= T_1^\dagger \\
R_5 &= C_3^\dagger R_1 C_3 & S_5 &= C_3^\dagger S_1 C_3 & T_5 &= C_3^\dagger T_1 C_3 \\
R_6 &= C_3 R_1^\dagger C_3^\dagger & S_6 &= C_3 S_1^\dagger C_3^\dagger & T_6 &= C_3 T_1^\dagger C_3^\dagger.
\end{aligned}
\end{equation}

From these relations, we can construct $H_0(\bm{k})$ using Eq.~\eqref{Hk} in terms of the parameters that comprise the matrices $E$, $R_1$, $S_1$, and $T_1$. With the exception of $\mu_0$, these parameters are then fit by minimizing the mean-square deviation between the band structure of $H_0(\bm{k})$ and our \textit{ab initio} calculation of the TaS$_2$ monolayer band structure. The computed values are displayed in Table \ref{tab:params}. The chemical potential $\mu_0$, in turn, is set at -50 meV to account for charge transfer between the monolayer and the substrate, as detailed in Sec.~\ref{model}.

\begin{table}[h]
\caption{\label{tab:params}%
Parameters used in the tight-binding model for 4Hb-TaS$_2$. The names of the terms follow the convention used in Tables II and III in \cite{Khodas2018}. The units are eV.
}
\begin{ruledtabular}
\begin{tabular}{cccccc}
$t_0$ & $t_1$ & $t_2$ & $t_{11}$ & $t_{12}$ & $t_{22}$\\
-0.1917 & 0.4057 & 0.4367 & 0.2739 & 0.3608 & -0.1845\\
$r_0$ & $r_1$ & $r_2$ & $r_{11}$ & $r_{12}$ & $r_{22}$\\
0.0409 & -0.069 & 0.0928 & -0.0066 & 0.1116 & 0\\
$u_0$ & $u_1$ & $u_2$ & $u_{11}$ & $u_{12}$ & $u_{22}$\\
0.0405 & -0.0324 & -0.0141 & 0.1205 & -0.0316 & -0.0778\\
$\epsilon_0$ & $\epsilon_1$ & $\epsilon_2$ & $\mu_0$ & $\lambda_{\text{SO}}$\\
1.6507 & 2.5703 & 2.5703 & -0.0500 & 0.1713\\
\end{tabular}
\end{ruledtabular}
\end{table}

\section{Definition of the Characteristic Angle}
\label{theta}

Here we calculate the angle $\theta_{\bm{k}}$, which in turn defines the topological charges of the nodes via its winding, as shown in Fig.~\ref{winding}.

First, we change the basis of only the hole block of the BdG spinor, as follows:
\begin{equation}
\tilde{\Psi}=\left[\begin{array}{cc}
\mathbb{I} & 0\\
0 & i\sigma_y
\end{array}\right]\left[\begin{array}{c}
\psi_{\bm{k}}\\
\psi_{-\bm{k}}^{\dagger}
\end{array}\right],
\end{equation}
where $i\sigma_y$ is the unitary part of the time reversal operator, acting as a Pauli matrix in spin and as the identity in orbitals. It is simple to show that this transformation, combined with the identity in Eq.~\eqref{DeltaIdentity}, requires that the matrix $\tilde{\Delta}$ in the new basis is Hermitian. This in turn means that the Hamiltonian in this new basis has no $\tau_y$ component, where $\tau_i$ refer to Pauli matrices in the particle-hole degree of freedom.

Since the Hamiltonian in the new basis, $\tilde{H}_{SC}(k)$, contains only $\tau_x$ and $\tau_z$ components, we can rotate it so that it contains only $\tau_x$ and $\tau_y$ instead - that is, it is entirely off-diagonal in the particle-hole degree of freedom:
\begin{equation}
e^{i\frac{\pi}{4}\tau_x}\tilde{H}_{SC}(\bm{k})e^{-i\frac{\pi}{4}\tau_x} =\left[\begin{array}{cc}
0 & Q_{\bm{k}} \\
Q^\dagger_{\bm{k}} & 0
\end{array}\right].
\end{equation}

The arguments of the eigenvalues of the matrix $Q$ must be smooth functions of $\bm{k}$. The sum of these 6 phases at each momentum $\bm{k}$ defines the characteristic angle 
\begin{equation}
\label{eq:thetak}
\theta_{\bm{k}} \equiv \text{arg}(\text{det}Q_{\bm{k}}).
\end{equation}
This total phase is plotted in Fig.~\ref{winding}. The winding of $\theta_{\bm{k}}$ around each node determines the charge of the node. Whenever two nodes are created or annihilated, the sum of their winding numbers must be zero.

\section{Angular Momentum Basis}
\label{basis}

To convert between the Cartesian basis used in most of this work and the angular momentum basis in which it is simplest to define the interaction terms, we use the $6\cross6$ unitary basis transformation matrix
\begin{equation}
\mathcal{U}^{\rm a} = \sigma_0 \otimes \left[\begin{array}{ccc}
1 & 0 & 0 \\
0 & \frac{1}{\sqrt{2}} & \frac{i}{\sqrt{2}} \\
0 & \frac{1}{\sqrt{2}} & -\frac{i}{\sqrt{2}} \\
\end{array}\right],
\end{equation}
where $\sigma_0$ denotes the identity matrix in the spin degree of freedom and the remaining $3\cross3$ matrix is in the orbital degree of freedom.

A $6\cross6$ matrix $M^{\rm a}$ (where we use the superscript $^{\rm a}$ to denote the angular momentum basis) is converted to the Cartesian basis via
\begin{equation}
M = \mathcal{U}^{{\rm a}\dagger} M^{\rm a} \mathcal{U}^{\rm a},
\end{equation}
while the 4-field interaction tensor $U^{\rm a}_{ijkl}$ transforms as
\begin{equation}
U_{ijkl} = \mathcal{U}^{{\rm a}\dagger}_{ie} \mathcal{U}^{{\rm a}\dagger}_{jf} U^{J,{\rm a}}_{efgh} \mathcal{U}^{\rm a}_{gk} \mathcal{U}^{\rm a}_{hl}.
\end{equation}
We perform this transformation to obtain $U_{ijkl}$ as it appears in Eq.~\eqref{Utensor}.

\section{Derivation of the RG Equation}
\label{derivation}

Here we derive Eqs.~\eqref{rgeq} and \eqref{nueq} via a generalized renormalization group process.

Starting with the interacting action in Eq.~\eqref{SU}, we perform a Hubbard-Stratonovich transformation to obtain the self-interaction term of the pairing matrix $\Delta$
\begin{equation}
S_{\Delta} = \Delta^\dagger_{ij} U^{-1}_{ijkl} \Delta_{kl},
\end{equation}
where $U^{-1}$ is obtained by treating $U_{ijkl}$ as a $36\cross36$ matrix, combining $ij$ into the first index and $kl$ into the second before inverting.

We integrate over a shell of energies between $\Lambda-d\Lambda$ and $\Lambda$, yielding the equality
\begin{widetext}
\begin{equation}
\label{shell}
\Delta^\dagger_{ij}[U^{-1}(\Lambda-d\Lambda) - U^{-1}(\Lambda)]_{ijkl} \Delta_{kl} = -T\sum_{\omega_n}{\sum_{b,\bm{k}}}'\text{ln det}
\left[\begin{array}{cc}
G_b^{-1}(\omega_n,\bm{k}) & \Delta\\
\Delta^\dagger & G_b^{*-1}(\omega_n,\bm{-k})
\end{array}\right],
\end{equation}
\end{widetext}
where $T$ is temperature and the sum ${\sum_{b,\bm{k}}}'$ is defined to be taken only over momenta $\bm{k}$ and band indices $b$ such that the energy of the band $E_b(\bm{k})$ of the diagonalized Hamiltonian $H_0(\bm{k})$ is within the shell (that is, $\Lambda-d\Lambda < |E_b(\bm{k})| < \Lambda$).

The $12\cross12$ matrix in the determinant is expressed in block form, where
\begin{equation}
G_b(\omega_n,\bm{k}) \equiv  \frac{1}{i\omega_n - E_b(\bm{k})} \ket{\psi_b}\bra{\psi_b}
\end{equation}
is the Green's function of $H_0(\bm{k})$ projected to the band index $b$. In the eigenbasis of $H_0(\bm{k})$, the only nonzero element of this $6\cross6$ matrix is the diagonal entry of band $b$. We work in the basis of spins and orbitals, so this matrix must be transformed appropriately using $\mathcal{U}(\bm{k})$, the unitary matrix with the eigenvectors of $H_0(\bm{k})$ as its columns.

Expanding the logarithm on the right side of Eq.~\eqref{shell} in the limit of small $\Delta$, we take only the linear terms and match the coefficients on both sides. The result is
\begin{flalign}
\label{expand}
[U^{-1}(\Lambda&-d\Lambda) - U^{-1}(\Lambda)]_{ijkl} \nonumber & \\
&= -T\sum_{\omega_n}{\sum_{b,\bm{k}}}'
(G_b^*)^{-1}_{jk}(\omega_n,\bm{-k}) (G_b)^{-1}_{li}(\omega_n,\bm{k}).
\end{flalign}

Evaluating the sum over Matsubara frequencies, the temperature cancels and we obtain a factor of $\frac{1}{E_b(\bm{k})}$, which within our shell sum is simply equal to $\frac{1}{\Lambda}$. This yields
\begin{flalign}
\label{expand}
[U^{-1}(\Lambda&-d\Lambda) - U^{-1}(\Lambda)]_{ijkl} \nonumber & \\
&= \frac{1}{\Lambda}{\sum_{b,\bm{k}}}'
\mathcal{U}^*_{ib}(\bm{k})
\mathcal{U}^*_{jb}(\bm{k})
\mathcal{U}_{kb}(\bm{k})
\mathcal{U}_{lb}(\bm{k}).
\end{flalign}

From here, we rewrite the sum as an integral and apply the basis transformation in Eq.~\eqref{transform4x4} to arrive at Eqs.~\eqref{rgeq} and \eqref{nueq}.

\bibliographystyle{aipnum4-2}
%

\end{document}